\RequirePackage[2020-02-02]{latexrelease}
\documentclass[times, twoside]{zHenriquesLab-StyleBioRxiv}

\usepackage{placeins}
\newcommand{\code}[1]{\texttt{#1}}
\usepackage{lineno}
\usepackage[utf8]{inputenc} 
\usepackage[T1]{fontenc}    
\usepackage{booktabs}       
\usepackage{amsfonts}       
\usepackage{nicefrac}       
\usepackage{microtype}      
\usepackage{xcolor}         
\usepackage{ulem}
\usepackage{color,soul}
\usepackage{placeins}
\usepackage{graphicx}
\usepackage{footmisc}
\usepackage{multirow}
\usepackage{amssymb}
\usepackage{enumitem}

\def\sensorium{\texttt{SENSORIUM}}

\leadauthor{Turishcheva*, Fahey*} 

\begin{document}


\title{The Dynamic Sensorium competition for predicting large-scale mouse visual cortex activity from videos}
\shorttitle{\sensorium~ 2023}
 


\author[1,*, \Letter]{Polina Turishcheva}
\author[2--6,*, \Letter]{Paul G. Fahey}
\author[1]{Michaela Vystrčilová}
\author[1]{Laura Hansel}
\author[2--6]{Rachel Froebe}
\author[2,3]{Kayla Ponder}
\author[1,4--6]{Yongrong Qiu}
\author[1,4--8]{Konstantin F. Willeke}
\author[1,7,8]{Mohammad Bashiri}
\author[2,3]{Eric Wang}
\author[2,3]{Zhiwei Ding}
\author[2--6,9*]{\\Andreas S. Tolias}
\author[1,2,3,7,8,*]{Fabian H. Sinz}
\author[1,10,*]{Alexander S. Ecker}

\affil[1]{Institute of Computer Science and Campus Institute Data Science, University of Göttingen, Germany}
\affil[2]{Department of Neuroscience, Baylor College of Medicine, Houston, TX, USA}
\affil[3]{Center for Neuroscience and Artificial Intelligence, Baylor College of Medicine, Houston, TX, USA}
\affil[4]{Department of Ophthalmology, Byers Eye Institute, Stanford University School of Medicine, Stanford, CA, US}
\affil[5]{Stanford Bio-X, Stanford University, Stanford, CA, US}
\affil[6]{Wu Tsai Neurosciences Institute, Stanford University, Stanford, CA, US}
\affil[7]{International Max Planck Research School for Intelligent Systems, University of Tübingen, Germany}
\affil[8]{Institute for Bioinformatics and Medical Informatics, University of Tübingen, Germany}
\affil[9]{Department of Electrical Engineering, Stanford University, Stanford, CA, US}
\affil[10]{Max Planck Institute for Dynamics and Self-Organization, Göttingen, Germany}

\affil[*]{Equal contributions}

\maketitle

\begin{abstract}
Understanding how biological visual systems process information is challenging due to the complex nonlinear relationship between neuronal responses and high-dimensional visual input. 
Artificial neural networks have already improved our understanding of this system by allowing computational neuroscientists to create predictive models and bridge biological and machine vision.
During the Sensorium 2022 competition, we introduced benchmarks for vision models with static input (i.e. images). 
However, animals operate and excel in dynamic environments, making it crucial to study and understand how the brain functions under these conditions. 
Moreover, many biological theories, such as predictive coding, suggest that previous input is crucial for current input processing. 
Currently, there is no standardized benchmark to identify state-of-the-art dynamic models of the mouse visual system. 
To address this gap, we propose the Sensorium 2023 Benchmark Competition with dynamic input (\url{https://www.sensorium-competition.net/}).
This competition includes the collection of a new large-scale dataset from the primary visual cortex of ten mice, containing responses from over 78,000 neurons to over 2 hours of dynamic stimuli per neuron.
Participants in the main benchmark track will compete to identify the best predictive models of neuronal responses for dynamic input (i.e. video).
We will also host a bonus track in which submission performance will be evaluated on out-of-domain input, using withheld neuronal responses to dynamic input stimuli whose statistics differ from the training set.  
Both tracks will offer behavioral data along with video stimuli. 
As before, we will provide code, tutorials, and strong pre-trained baseline models to encourage participation. 
We hope this competition will continue to strengthen the accompanying Sensorium benchmarks collection as a standard tool to measure progress in large-scale neural system identification models of the entire mouse visual hierarchy and beyond.

\end{abstract}

\begin{corrauthor}
turishcheva\at cs.uni-goettingen.de; pgfahey\at stanford.edu; ecker\at cs.uni-goettingen.de
\end{corrauthor}

\subsection*{Keywords}
mouse visual cortex, system identification, neural prediction, dynamic stimulus

\begin{figure*}[ht!]
    \begin{minipage}[t]{0.95\linewidth}
        \vspace{0 pt}
        \includegraphics[trim=0 0 0 0, clip,width=\linewidth]{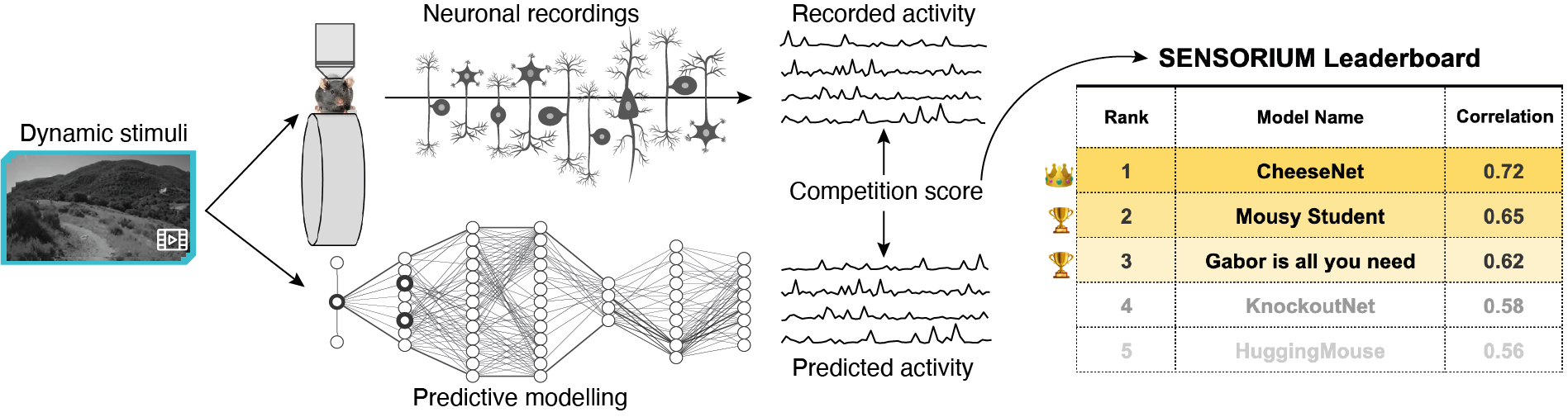}
    \end{minipage}\hfill
    \begin{minipage}[t]{.95\linewidth}
        \vspace{20pt}
        \caption{\textbf{A schematic illustration of the \sensorium~ competition.}
             We will provide large-scale datasets of neuronal activity in the primary visual cortex of mice. Participants of the competition will train models on pairs of natural image stimuli and recorded neuronal activity, in search for the best neural predictive model. 
        }\label{fig:competition_overview}
    \end{minipage}
\end{figure*}

\section*{Introduction}

Understanding how the visual system processes visual information has been a longstanding goal of neuroscience.
Neural system identification, the development of accurate predictive models of neural population activity in response to arbitrary input, is a powerful approach to develop our understanding on a quantitative, testable, and reproducible basis.
Systems neuroscience has used a variety of modeling approaches to study the visual cortex in the past, including linear-nonlinear (LN) models \citep{simoncelli2004characterization, Jones1987-sn,Heeger1992-ig, Heeger1992-xx}, energy models \citep{Adelson1985-re}, subunit models \citep{liu2017inference, rust2005spatiotemporal, touryan2005spatial, vintch2015convolutional}, Bayesian models \citep{walker2020neural, george2005hierarchical}, redundancy reduction models \citep{perrone2015redundancy}, and predictive coding models \citep{marques2018functional}. 
Deep learning has significantly advanced the performance of predictive models, particularly with the introduction of convolutional neural networks (CNNs) trained on image recognition tasks \citep{Yamins2014,Cadieu2014,Cadena2019} or trained end-to-end on predicting neural responses \citep{Cadena2019,Antolik2016,batty2017multilayer,McIntosh2016,Klindt2017,Kindel2017,burg2021learning, Lurz2020-ua, Bashiri2021-or,Zhang2018-cs,Cowley2020neurips,Ecker2018, sinz2018stimulus, Walker2019-oq,Franke2022, Wang2023.03.21.533548, Fu2023.03.13.532473, ding2023bipartite}. 
More recently, transformer-based architectures have also shown strong performance in predicting neural responses \citep{li2023v1t}. 

In some cases, predictive models may be engineered with specific constraints in order to draw insight from interpretable internal parameters.  
On the other hand, even ``black-box'' models can still provide important scientific utility.  
For example, high-performing, data-driven models allow unbiased exploration of large stimulus spaces \textit{in silico} that would otherwise be prohibitively costly with biological experiments, yielding novel insights about the visual system that are evaluated by selective verification by systems neuroscientists \textit {in vivo}~\citep{Walker2019-oq,Ponce2019-yn,Bashivan2019,Franke2022,hoefling2022chromatic,Fu2023.03.13.532473,ding2023bipartite, ustyuzhaninov2022digital,Wang2023.03.21.533548}.
Additionally, another research focus could be to develop models that generalize well from the training domain (e.g. natural movies) to novel out-of-domain stimuli \citep{ren2023well}. 
Such models can also dramatically extend the variety of questions that can be asked of the same dataset by characterizing classical vision tuning properties (e.g. orientation tuning and receptive field location) or novel hypothesis-driven tuning that may be costly or impossible to characterize \textit{in vivo} \citep{Wang2023.03.21.533548,Ding2023-funconn,ding2023bipartite,Fu2023.03.13.532473}.
Thus, improving predictive performance of these models opens up new avenues for important neuroscientific inquiry.

Standardized large-scale benchmarks are one important approach to steadily accumulate improvements in predictive models, through constructive competition between models compared on equal ground \citep{dean2018new}.  
Several neuroscience benchmarks already exist, including \textit{Brain-Score} \citep{Schrimpf2018,Schrimpf2020-hd}, \textit{Neural Latents '21} \citep{pei2021neural}, \textit{Algonauts} \citep{Cichy2019-re, Cichy2021-lr, gifford2023algonauts} and \textit{Sensorium 2022} \citep{sensorium_whitepaper}.  
There are also several recent large datasets that have been released as high-throughput recording methodologies become more available, including the MICrONS calcium imaging dataset \citep{microns2021functional} and calcium imaging and Neuropixel datasets from the Allen Brain Observatory \citep{de2020large, Siegle2021-en}
However, these large public datasets typically lack the private test set and benchmark infrastructure for third party evaluation of performance metrics on withheld test data. 

Importantly, the majority of the above models, competitions, and datasets focus on predicting responses to static stimuli, typically with relatively long presentation times (i.e. hundreds of milliseconds). 
While this approach has yielded important insights into the spatial preferences of neural populations, understanding how visual neurons process spatiotemporal information is crucial, because real-life visual stimuli are dynamic.
Animals need to be able to accurately and quickly detect and respond to external elements in their environment (e.g. when tracking prey or avoiding a predator), as well as correctly estimate their own motion.
Thus, further developing and assessing the performance of models designed for neural predictions over time \citep{sinz2018stimulus, Wang2023.03.21.533548,zheng2021unraveling, batty2017multilayer, McIntosh2016} is important. 
However, \emph{the field currently lacks a large-scale benchmark for models predicting single-cell responses to dynamic (movie) stimuli}.

\begin{figure*}[ht!]
    \begin{minipage}[t]{1\linewidth}
        \vspace{0 pt}
        \centering
        \includegraphics[trim=0 0 0 0, clip,width=\linewidth]{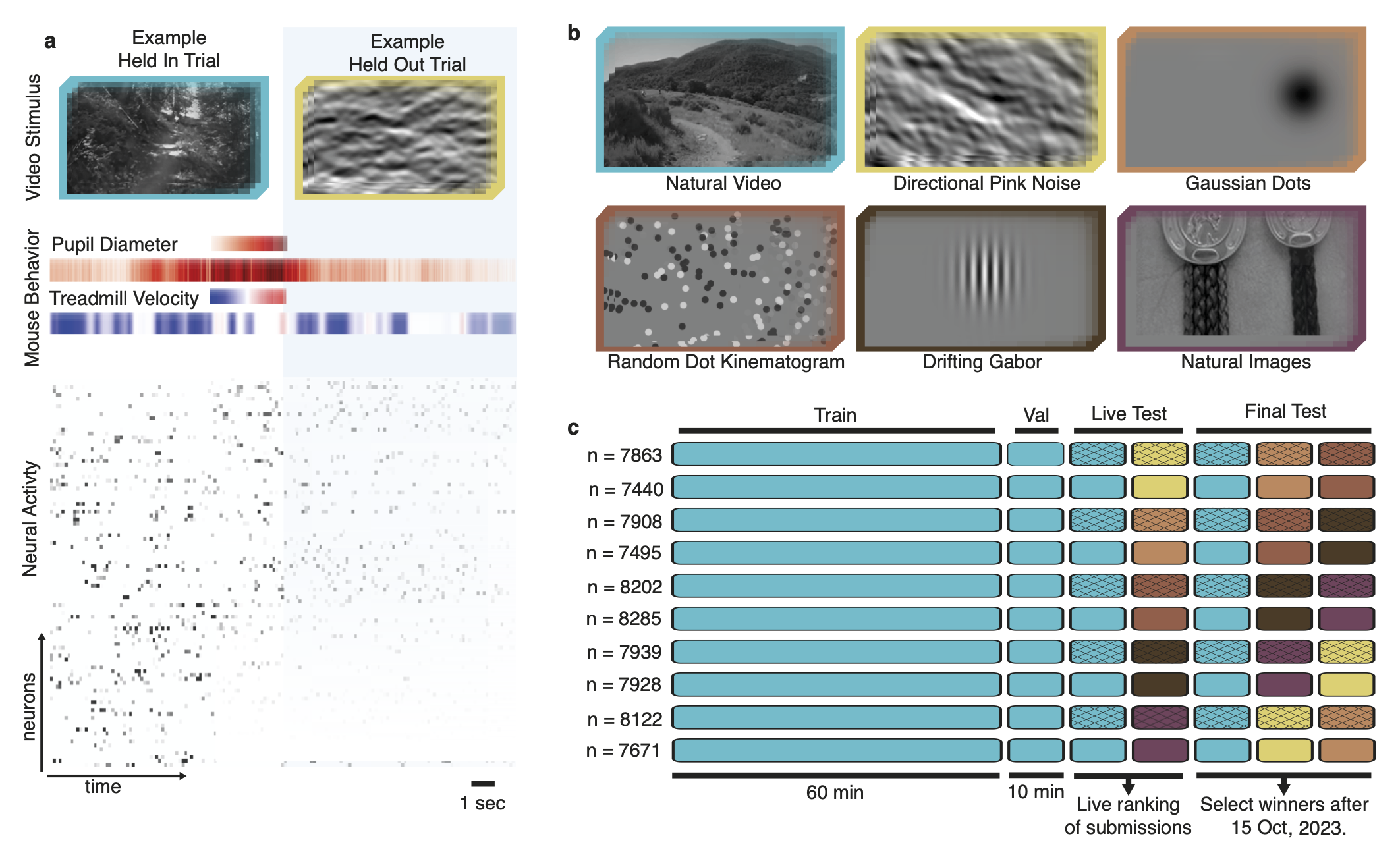}
    \end{minipage}\hfill
    \begin{minipage}[t]{0.95\linewidth}
        \vspace{5 pt}
        \caption{\textbf{Overview of the data.}
             \textbf{a}, Example stimulus frames, behavior (pupil position not depicted) and neural activity. 
             \textbf{b}, Representative frames from natural video and five OOD stimuli. 
             \textbf{c}, Stimulus composition (color) and availability for ten scans in ten animals.  
             The crossed elements were used for live and final test sets in the competition evaluation.
             For non-crossed elements all responses are available. $n$ is number of neurons per scan.}
   \label{fig:data_overview}
    \end{minipage}
\end{figure*}

To address this gap, we propose the \sensorium\ 2023 competition, aimed at fostering the development of more accurate predictive dynamic models of the mouse visual cortex.
These predictive dynamic models take as input video stimuli and/or behavioral variables, and as output predict video-rate responses of single neurons  (\cref{fig:competition_overview}).
We designed and collected a large-scale dataset for this competition, including ten scans from the primary visual cortex of ten mice.
In total, the dataset contains responses from 78,853 neurons to a diverse set of videos from various domains, along with behavioral measurements  (\cref{fig:data_overview}). 
The main track will focus on predicting neuronal activity in response to natural videos, with participants encouraged to use behavioral data to enhance their predictions.
To test how well the models generalize, a bonus track will evaluate model performance on five out-of-domain stimuli not included in the training set, including parametric stimuli that have been used to characterize classical visual tuning properties.
We also provide a starting kit to lower the barrier for entry, with tutorials, code for training baseline models, and APIs for data loading and submission.
This competition is part of an ongoing series of \sensorium\ competitions for benchmarking predictive models of neuronal responses with the hope that it facilitates our understanding of the computations carried out by visual sensory neurons. 

\section*{\sensorium\ competition overview}
\label{sec:Competition overview}

The goal of the \sensorium~2023 competition is to identify accurate predictive dynamic models of mouse visual cortex.
Participants are provided with training data in the form of videos that were shown to the mouse, and the resulting recorded neuronal responses and behavioral variables, all of which were recorded for this purpose and will be made public for the first time as part of the competition \footnote[1]{\url{http://sensorium-competition.net/}}.
Participants are then tasked with creating models that predict a test set of withheld neuronal responses from the corresponding video stimuli and behavioral variables.
Submissions to the main track are evaluated on a test set of natural video stimuli of the same type present in the training set (i.e. in-domain performance).
Submissions to the bonus track are evaluated on a test set of stimulus types not present in the training set (i.e., out-of-domain performance), including static natural images, random dot kinematograms, drifting gabors, gaussian dots, and directional pink noise, as in ~\citet{Wang2023.03.21.533548}.
For the neurons evaluated as a part of the competition no OOD responses are provided.

The test set trials are divided into two exclusive groups: \textit{live} and \textit{final test}.
Performance metrics computed on the \textit{live test} trials will be used to maintain a public leaderboard throughout the submission period, while the performance metrics on the \textit{final test} trials will be used to identify the winning entries, and will only be revealed after the submission period has ended (\cref{fig:data_overview}d).  
By separating the \textit{live test} and \textit{final test} set performance metrics, we are able to provide feedback from the \textit{live test} set to participants wishing to submit updated predictions over the course of the competition (up to one submission per day), while avoiding overfitting for the \textit{final test} set over multiple submissions.
In both cases, the withheld competition test set responses will not be (and have never been) publicly released.


To make the competition accessible for both computational neuroscientists and machine learning practitioners, we will release a \textit{starting kit} that contains the complete code to fit our baseline models as well as explore the full dataset.\footnote[2]{\url{https://github.com/ecker-lab/sensorium_2023/}}

\section*{Data}
\label{sec:Data}
We recorded data with the goal of comparing models that predict neuronal activity in response to dynamic movies. 
We also include behavioral variables in our dataset as a common proxy of modulatory effects of neuronal responses \citep{niell2010modulation, reimer2014pupil}. 
Thus, in generic terms, neural predictive models capture neural responses $\mathbf{r} \in \mathbb{R}^{n\times t}$ of $n$ neurons for $t$ timepoints as a function $\mathbf{f}_\theta(\mathbf{x}, \mathbf{b})$ of both natural movie stimuli $\mathbf{x} \in \mathbb{R}^{w\times h\times t}$, where $w$ and $h$ are video width and height, and behavioral variables $\mathbf{b} \in \mathbb{R}^{k\times t}$, where $k$ is the types of behavior ($k=4$, see below).  
In the following paragraphs, we provide a short description of each one of these quantities.

\vspace{-5 pt}
\subsection*{Movie stimuli}
We sampled natural dynamic stimuli from cinematic movies and the Sports-1M dataset \citep{Karpathy2014-gu}, as described in \citep{microns2021functional}.
Five additional out of domain (OOD) stimulus types, including natural images from ImageNet \citep{Russakovsky2015-xi, Walker2019-oq}, flashing Gaussian dots, random dot kinematograms \citep{Morrone2000-ov}, directional pink noise \citep{microns2021functional}, and drifting Gabors \citep{Petkov2007-kv} were also included in the stimulus, in line with earlier work \citep{Wang2023.03.21.533548}.
Stimuli were converted to grayscale and presented to mice in $\sim8-11$ second clips at 30 Hz (\cref{fig:data_overview}b).

\vspace{-5 pt}
\subsection*{Neuronal responses}
Using a wide-field two-photon microscope \citep{sofroniew2016large}, we recorded the responses of excitatory neurons at 8 Hz in layers 2--5 of the right primary visual cortex in awake, head-fixed, behaving mice using calcium imaging. 
Neuronal activity was extracted as described previously \citep{Wang2023.03.21.533548} and resampled at 30 Hz to be at the same frame rate as the visual stimuli (\cref{fig:data_overview}a). 
We will also release the anatomical coordinates of the recorded neurons. 

\vspace{-5 pt}
\subsection*{Behavioral variables}
We provide measurements of four behavioral variables: \emph{locomotion speed}, which is recorded from a cylindrical treadmill at 100 Hz and resampled to 30 Hz, and \emph{pupil size, horizontal and vertical pupil center position}, which are extracted from tracked eye camera video at 20 Hz and resampled to 30 Hz. 

\vspace{-5 pt}
\subsection*{Dataset}
Our complete corpus of data comprises ten recordings in ten animals, which in total contain the neuronal activity of 78,853 neurons to a total of $\sim$ 1200 minutes of dynamic stimuli over the dataset, with $\sim$ 120 minutes per recording (\cref{fig:data_overview}c).
None of the ten recordings have been published before, and were released as part of this competition explicitly for this purpose.
Five were released on the first day of the competition, but accidentally included responses for live and final test sets, and were thus released in their entirety as pretraining data.
An additional five scans were collected for competition evaluation, and were added to the release with live and final test set data withheld. 

Each animal recording consists of 4 components (\cref{fig:data_overview}c):  
\begin{itemize}[noitemsep,nolistsep,leftmargin=*]
    \item \textbf{Training set:} 60 minutes of natural movies, one repeat each (60 minutes total).
    \item \textbf{Validation set:} 1 minute of natural movies, ten repeats each (10 minutes total). 
    \item \textbf{Live test set:} 1 minute of natural movies and 1 minute of OOD stimuli, ten repeats each (20 minutes total).
    Each OOD stimulus type is represented once in the live test set across the five recordings.
    \item \textbf{Final test set:} 1 minute of natural movies and 2 minutes of OOD stimuli, ten repeats each (30 minutes total).
    Each OOD stimulus type is represented twice in the final test set across the five recordings. 
\end{itemize}
For the first five mice the responses for all components are released, while for the mice used for competition evaluation only train and validation responses are available publicly.
For the training set and validation set, the stimulus frames, neuronal responses, and behavioral variables are released for model training and evaluation by the participants, and are not included in the competition performance metrics.  
Please note that train and validation sets for the mice used for evaluation only contain natural movies and not the OOD stimuli.


\vspace{-5 pt}
\subsection*{Data availability}
The complete corpus of data is available for download \footnote[1]{\url{http://sensorium-competition.net/}}. 
To decrease the requirement for local storage, the competition dataset is available through Deep Lake \citep{hambardzumyan2022deep} in a convenient format for use with standard model-fitting techniques, allowing caching data subsets for training.

\begin{table*}[t]
\centering
\begin{tabular}{lllllllll}
\hline
                            &              & \multicolumn{7}{c}{Baseline performance on held out test data}    \\ 
\hline
                            &              & \multicolumn{2}{c}{Live test set}&  & \multicolumn{2}{c}{Final test set}\\ 
\cline{3-4}\cline{6-7}
Competition track           &              & \multicolumn{1}{c}{\begin{tabular}[c]{@{}c@{}}Single Trial\\ Correlation\end{tabular}} & \multicolumn{1}{c}{\begin{tabular}[c]{@{}c@{}}Correlation\\ to Average\end{tabular}}  &  & \multicolumn{1}{c}{\begin{tabular}[c]{@{}c@{}}Single Trial\\ Correlation\end{tabular}} & \multicolumn{1}{c}{\begin{tabular}[c]{@{}c@{}}Correlation\\ to Average\end{tabular}}  \\ 
\hline
\multirow{2}{*}{Main Track} 
& Ensembled Baseline & \multicolumn{1}{c}{.206} & \multicolumn{1}{c}{.381} &  & \multicolumn{1}{c}{\textbf{.197}} & \multicolumn{1}{c}{.371} \\
& 3D Factorized Baseline & \multicolumn{1}{c}{.177}& \multicolumn{1}{c}{.337}&  & \multicolumn{1}{c}{.164}& \multicolumn{1}{c}{.321}\\
& GRU Baseline  & \multicolumn{1}{c}{.108}& \multicolumn{1}{c}{.207}&  & \multicolumn{1}{c}{.106}& \multicolumn{1}{c}{.207}\\
                            &              & & & &  & & & \\
\multirow{2}{*}{Bonus Track} 
& Ensembled Baseline & \multicolumn{1}{c}{.128} & \multicolumn{1}{c}{.234} &  & \multicolumn{1}{c}{\textbf{.129}} & \multicolumn{1}{c}{.241} \\
& 3D Factorized Baseline & \multicolumn{1}{c}{.112} & \multicolumn{1}{c}{.204} &  & \multicolumn{1}{c}{.121} & \multicolumn{1}{c}{.223} \\
& GRU Baseline  & \multicolumn{1}{c}{.061} & \multicolumn{1}{c}{.106} &  & \multicolumn{1}{c}{.059} & \multicolumn{1}{c}{.106}  \\
\hline
\end{tabular}
\caption{
        \textbf{Performance of the baseline models on both competition tracks.}
        Three baseline scores are provided for each competition track. 
        The minimum performance for winning entries is indicated in bold.
        }
\label{table:T1}
\end{table*}


\section*{Baseline models}
\label{sec:Baseline}
\sensorium~2023 is accompanied by three model baselines (\cref{table:T1}):
\begin{itemize}[noitemsep,nolistsep,leftmargin=*]
    \item \textbf{GRU baseline} is a dynamic model with a 2D CNN core and gated recurrent unit (GRU) inspired by earlier work \citep{sinz2018stimulus}, but replacing the factorized readouts with more recently developed Gaussian readouts as in \citep{Lurz2020-ua}. 
    Conceptually, the 2D core transforms each frame of the video visual stimulus into a latent space, and the GRU persists certain elements from this latent space through time.
    The Gaussian readout learns the spatial preference of each neuron in visual space (``receptive field''), the position at which a vector from the latent space is extracted.
    This latent vector is convolved by a vector of weights learned per neuron (``embedding'') to predict the activity of the neuron at a specified time point. 
    \item \textbf{Factorized baseline} is a dynamic model with a 3D factorized convolution core and Gaussian readouts inspired by earlier work \citep{hoefling2022chromatic}. 
    In contrast with the GRU baseline, where the 2D CNN core does not interact with the temporal component, the factorized core learns both spatial and temporal components at each layer. 
    This allows the model to transform both spatial and temporal components iteratively as the dimensionality of the latent space changes with increasing channels across layers.
    \item \textbf{Ensembled baseline} is simply an ensembled version of the above factorized baseline over 14 models.
    Ensembling is a well-known tool to improve the model performance in benchmark competitions \citep{allenzhu2023understanding}.
    In order to focus the results of the competition on novel architectures and training methods beyond simple ensembling, only entries outperforming the ensembled baseline will be candidates for competition winners.
\end{itemize}

\section*{Competition Evaluation}
Similar to \sensorium~2022, for each submission we will compute and report two metrics: 
\begin{itemize}[noitemsep,nolistsep,leftmargin=*]
    \item \textbf{Single-trial correlation} on the natural video final test set will be used to determine competition winners for the main track. 
    We will also compute the single-trial correlation metric for each of the five OOD stimulus test sets separately, and the mean single-trial correlation across all five OOD final test sets will be used to determine the competition winner for the bonus track.
    \item \textbf{Correlation to average} is also calculated for research purposes.  
    This metric is more robust to noise due to averaging ground truth data over repeats, but it does not measure how well a model accounts for stimulus-independent variability caused by behavioral fluctuations.

\end{itemize}

For all metrics, the first 50 frames of the prediction and neuronal responses will be discarded before computing.  
This is to allow a ``burn-in'' period for dynamic models that rely on history to reach maximum performance.  
For more details and equations, please see Methods.

\section*{Discussion}
Here, we introduced the \sensorium~2023 competition for finding the best predictive model for neuronal responses in mouse primary visual cortex to dynamic stimuli.
This competition is the second in a series, and shares much of its structure with preceding year's competition, \sensorium~2022.
Similar to last year, we have included a starting kit with baseline model tutorials, in order to continue supporting accessibility for both neuroscientists and machine learning experts interested in participating. 
We also once again collected a dedicated large-scale dataset, including an estimated 246\% increase in unique neuron-hours above the preceding year.  
Importantly, we made several major changes in this iteration, including moving from static to dynamic stimuli, adding out-of-domain performance in the bonus track, and including behavior in both tracks. 
These changes pose new technical challenges and broaden the variety of scientific questions to work on.
    
The \sensorium\ 2023 challenge differs from existing benchmarks in that it is the only benchmark for predicting single-cell responses to dynamic natural movie stimuli. 
The \textit{Brain-Score} benchmark \citep{Schrimpf2018,Schrimpf2020-hd} recently added a dynamic component, asking models to predict the temporal evolution of neural activity, but it still focuses on static images as stimuli. 
In addition, its scientific goal is not to benchmark predictive models, but to evaluate how well task-pretrained computer vision models match the neural representations along the primate ventral stream. 
The \textit{Neural Latents Benchmark '21} \citep{pei2021neural} tests models of neural population activity, but focuses on dimensionality reduction and extracting a small set of latent variables from high-dimensional neural population activity, not necessarily in response to visual stimuli. 
The \textit{Algonauts challenge} is similar in spirit to last year's \sensorium\ 2022 and focuses on predictive models in response to natural images \citep{Cichy2019-re,gifford2023algonauts} or natural video \citep{Cichy2021-lr}, but tests models of functional magnetic resonance imaging (fMRI) in human visual cortex, as opposed to single-cell responses in mouse as in \sensorium.

This competition also departs from \sensorium~2022 in that both tracks now include behavioral measurements as model inputs. 
One key issue in assessing model performance is the fact that neural responses are noisy -- repeated presentation of the same stimulus does not produce identical responses. 
This question has been addressed by numerous authors, but no clear consensus has emerged \citep{roddey2000assessing, hsu2004quantifying, Haefner2008, Schoppe2016, pasupathy2001shape}. 
The usual solution is to attempt to estimate the trial-to-trial variability through the use of repeated stimulus presentations, and then estimate a noise-corrected version of the explained variance \citep[See][for an in-depth discussion and evaluation of existing metrics as well as a proposal of an asymptoticaly unbiased estimator]{Pospisil2020}. 
However, not everything determining neural responses is under experimental control.
For example, the freely varying behavioral state of the animal modulates neuronal responses, and by including behavioral variables as predictors we can increase the model predictive performance. 
Yet in consequence, we lose the ability to estimate the ``noise'' level, because every trial is now a unique combination of behavior and stimulus.
As a result, there is no way to determine the maximum achievable performance of a model without additional assumptions, and thus existing approaches for addressing unexplainable trial-to-trial fluctuations are not applicable. 
For this reason we opted to use the simplest possible measure of performance: the correlation coefficient between model prediction and observed response on a single-trial basis. 
While this metric serves our primary purpose of comparing models, it lacks the desirable property of assigning a perfect model a correlation of~1. 
Whether and how it is possible to obtain performance estimates with non-vacuous upper bounds once behavioral variables are included as model predictors is an open research question for future work.

We plan to continue running the family of \sensorium\ competitions with regular dataset releases and challenges, which will persist as benchmarks once the competition has ended. 
Our hope is these competitions and datasets are not only a technical resource, but also a basis for community formation around developing and testing models.
We expect that encouraging discussion around predictive modeling between machine learning practitioners and computational neuroscientists will create opportunities to exchange ideas and benefit from each other's expertise. 

\section*{Acknowledgments}
FHS is supported by the Carl-Zeiss-Stiftung and acknowledges the support of the DFG Cluster of Excellence “Machine Learning – New Perspectives for Science”, EXC 2064/1, project number 390727645 as well as the German Federal Ministry of Education and Research (BMBF) via the Collaborative Research in Computational Neuroscience (CRCNS) (FKZ 01GQ2107). 
This work was supported by an AWS Machine Learning research award to FHS. 
MB and KW were supported by the International Max Planck Research School for Intelligent Systems. 
This project has received funding from the European Research Council (ERC) under the European Union’s Horizon Europe research and innovation programme (Grant agreement No. 101041669). 

The project received funding by the Deutsche Forschungsgemeinschaft (DFG, German Research Foundation) via Project-ID 454648639 (SFB 1528), Project-ID 432680300 (SFB 1456) and Project-ID 276693517 (SFB 1233).

This research was supported by National Institutes of Health (NIH) via National Eye Insitute (NEI) grant RO1-EY026927, NEI grant T32-EY002520, National Institute of Mental Health (NIMH) and National Institute of Neurological Disorders and Stroke (NINDS) grant U19-MH114830, NINDS grant U01-NS113294, and NIMH grants RF1-MH126883 and RF1-MH130416.  
This research was also supported by National Science Foundation (NSF) NeuroNex grant 1707400. The content is solely the responsibility of the authors and does not necessarily represent the official views of the NIH, NEI, NIMH, NINDS, or NSF.

This research was also supported by the Intelligence Advanced Research Projects Activity (IARPA) via Department of Interior/Interior Business Center (DoI/IBC) contract no. D16PC00003, and with funding from the Defense Advanced Research Projects Agency (DARPA), Contract No. N66001-19-C-4020. 
The US Government is authorized to reproduce and distribute reprints for governmental purposes notwithstanding any copyright annotation thereon. 
The views and conclusions contained herein are those of the authors and should not be interpreted as necessarily representing the official policies or endorsements, either expressed or implied, of IARPA, DoI/IBC, DARPA, or the US Government.

\begin{contributions}

\textbf{PT}: Conceptualization, Formal Analysis, Investigation, Methodology, Project Administration, Software, Data Curation, Validation, Visualization, Writing - Original Draft, Writing - Review and Editing
\textbf{PGF}: Conceptualization, Data Curation, Methodology, Formal Analysis, Investigation, Methodology, Project Administration, Software, Validation, Visualization, Writing - Original Draft, Writing - Review and Editing
\textbf{LH}: Software, Writing - Original Draft, Writing - Review and Editing
\textbf{RF}: Investigation, Data Curation
\textbf{KP}: Investigation
\textbf{MV}: Software
\textbf{KFW}: Conceptualization, Methodology
\textbf{MB}: Data Curation
\textbf{EW}: Conceptualization, Methodology
\textbf{ZD}: Investigation
\textbf{AST}: Conceptualization, Methodology, Funding Acquisition, Supervision, Writing - Review and Editing
\textbf{FHS}: Conceptualization, Methodology, Funding Acquisition, Supervision, Writing - Review and Editing
\textbf{ASE}: Conceptualization, Methodology, Funding Acquisition, Supervision, Writing - Review and Editing
\end{contributions}


\section*{Materials and Methods}
\subsection*{Neurophysiological experiments} 
All procedures were approved by the Institutional Animal Care and Use Committee of Baylor College of Medicine. 
Ten mice (Mus musculus, 4 females, 6 males, P78--146 on day of first scan) expressing GCaMP6s in excitatory neurons via Slc17a7-Cre and Ai162 transgenic lines (recommended and generously shared by Hongkui Zeng at Allen Institute for Brain Science; JAX stock 023527 and 031562, respectively) were anesthetized and a 4 mm craniotomy was made over the visual cortex of the right hemisphere as described previously \citep{reimer2014pupil, Froudarakis2014-lx}. 

Mice were head-mounted above a cylindrical treadmill and calcium imaging was performed using Chameleon Ti-Sapphire laser (Coherent) tuned to 920\,nm and a large field of view mesoscope \citep{sofroniew2016large} equipped with a custom objective (excitation NA 0.6, collection NA 1.0,  21\,mm focal length). 
Laser power after the objective was increased exponentially as a function of depth from the surface according to:

\begin{equation}
    P = P_0 \times e^{(z/L_z)}
\end{equation}

Here P is the laser power used at target depth z, P0 is the power used at the surface (not exceeding 21\,mW), and $L_z$ is the depth constant (220\,μm).  
The greatest laser output of 94\,mW was used at approximately 425\,μm from the surface, with most scans not requiring more than 80\,mW at similar depths.   

The craniotomy window was leveled with regards to the objective with six degrees of freedom.  
Pixel-wise responses from an ROI spanning the cortical window (3600 $\times$ 4000\,μm, 0.2\,px/μm, approx. 200\,μm from surface, 2.47\,Hz) to drifting bar stimuli were used to generate a sign map for delineating visual areas \citep{Garrett2014-ki}.  
Area boundaries on the sign map were manually annotated.  
Our target imaging site was a 630 $\times$ 630\,µm ROI within the boundaries of primary visual cortex (VISp, Supp.~\cref{fig:imaging_site_selection}).  

The released scans contained 10 planes, with 25 μm interplane distance in depth, and were collected at 7.98 Hz.  Each plane is 630 $\times$ 630 μm (252 $\times$ 252 pixels, 0.4\,px/μm).  
The most superficial plane in each volume was approximately 200\,μm from the surface.  
This 25\,μm sampling in z was designed to reduce the number of redundant masks arising from multiple adjacent planes intersecting with the footprint of a single neuron.   

Movie of the animal's eye and face was captured throughout the experiment.  
A hot mirror (Thorlabs FM02) positioned between the animal's left eye and the stimulus monitor was used to reflect an IR image onto a camera (Genie Nano C1920M, Teledyne Dalsa) without obscuring the visual stimulus.  
The position of the mirror and camera were manually calibrated per session and focused on the pupil.  
Field of view was manually cropped for each session to contain the left eye in its entirety, ranging from 214--308 pixels height $\times$ 250--331 pixels width at ca. 20\,Hz.
Frame times were time stamped in the behavioral clock for alignment to the stimulus and scan frame times. 
Video was compressed using Labview’s MJPEG codec with quality constant of 600 and stored in an AVI file.

Light diffusing from the laser during scanning through the pupil was used to capture pupil diameter and eye movements.  
A DeepLabCut model \citep{Mathis2018-ti} was trained as previously described \citep{sensorium_whitepaper} on 17 manually labeled samples from 11 animals to label each frame of the compressed eye video (intraframe only H.264 compression, CRF:17) with 8 eyelid points and 8 pupil points at cardinal and intercardinal positions.  
Pupil points with likelihood >0.9 (all 8 in 55-99\% of frames) were fit with the smallest enclosing circle, and the radius and center of this circle was extracted.  
Frames with < 3 pupil points with likelihood >0.9 (<0.9\% frames per scan), or producing a circle fit with outlier > 5.5 standard deviations from the mean in any of the three parameters (center x, center y, radius, <0.3\% frames per scan) were discarded (total <0.9\% frames per scan).  
Gaps in behavior were replaced by linear interpolations over the whole session, if there were more than 2 frames with gaps, then the video is removed. 
(We removed  $\sim$2\% of the videos, 155 out of  7280, where one video was rejected due to signal synchronization issues during resampling). 

The mouse was head-restrained during imaging but could walk on a treadmill. 
Rostro-caudal treadmill movement was measured using a rotary optical encoder (Accu-Coder 15T-01SF-2000NV1ROC-F03-S1) with a resolution of 8000 pulses per revolution, and was recorded at approx. 100.2,Hz in order to extract locomotion velocity. 

\subsection*{Visual stimulation} 

Visual stimuli were presented with Psychtoolbox 3 in MATLAB \citep{Brainard1997-ed, Kleiner2007-ik, Pelli1997-vv} to the left eye with a 31.8 $\times$ 56.5\,cm (height $\times$ width) monitor (ASUS PB258Q) with a resolution of 1080$\times$1920 pixels positioned 15\,cm away from the eye.  
When the monitor is centered on and perpendicular to the surface of the eye at the closest point, this corresponds to a visual angle of 3.8\,°/cm at the nearest point and 0.7\,°/cm at the most remote corner of the monitor.  
As the craniotomy coverslip placement during surgery and the resulting mouse positioning relative to the objective is optimized for imaging quality and stability, uncontrolled variance in animal skull position relative to the washer used for head-mounting was compensated with tailored monitor positioning on a six dimensional monitor arm. 
The pitch of the monitor was kept in the vertical position for all animals, while the roll was visually matched to the roll of the animal’s head beneath the headbar by the experimenter. 
In order to optimize the translational monitor position for centered visual cortex stimulation with respect to the imaging field of view, we used a dot stimulus with a bright background (maximum pixel intensity) and a single dark square dot (minimum pixel intensity).  
Dot locations were randomly ordered from a 10 $\times$ 10 grid tiling a central square (approx. 90° width and height) with 10 repetitions of 200 ms presentation at each location.  
The final monitor position for each animal was chosen in order to center the population receptive field of the scan field ROI on the monitor, with the yaw of the monitor visually matched to be perpendicular to and 15 cm from the nearest surface of the eye at that position.

\underline{\textit{Natural Movies}}: 
Natural movies from the ``cinematic'' and ``Sports-1M'' \citep{Karpathy2014-gu} classes were drawn from the library described in \citep{microns2021functional}.  
Each scan contained 360 movies shown one time and 18 movies shown ten times, in both cases drawn equally from the cinematic and Sports-1M classes. 
Five sets of natural movies were prepared, with each movie unique to its respective set, and each set of movies shown in two scans.

\underline{\textit{Spatiotemporal Gabors}}: 
Spatiotemporal gabor movies were presented as described in \citep{Petkov2007-kv,Wang2023.03.21.533548}, but with different parameters as described below.
For six scans containing spatiotemporal gabors, 72 movies (8 directions $\times$ 3 spatial frequencies $\times$ 3 temporal frequencies) were shown ten times per scan.
Gabor spatial frequencies corresponded to wavelengths of 0.05, 0.1, and 0.2 (fraction of monitor width).
Gabor temporal frequencies corresponded to gabor velocities of 0.1, 0.2, and 0.3 (fraction of monitor width per second), in the direction perpendicular to the gabor orientation.  
Gabor spatial envelope was located in the center of the monitor, with a standard deviation of 0.08 (fraction monitor width, approx. 17 degrees). 
Each gabor movie was 833 ms in duration, and movies were randomly assorted into 6 sequences of 12 conditions each, for a total of 10 seconds per sequence. 
Because the stimulus was parametrically constructed, the same movies are shown in each of the six scans.  
Three sets of gabor movies that differ in sequence membership and order were prepared, and each set of movies was shown in two scans. 

\underline{\textit{Directional Pink Noise}}:  Directional pink noise was generated as described in \citep{microns2021functional}.  
For six scans with directional pink noise stimuli, six movie sequences were shown time times per scan.
Each movie sequence was generated from a unique random seed, which determined the underlying pink noise pattern and also the order of 12 equally spaced directional subtrials, with a spatial orientation bias perpendicular to the direction of motion.  
Each directional subtrial lasted 900 ms, for a total of 10.8 seconds per sequence.
Three sets of directional pink noise movie sequences were prepared, with each sequence unique to its respective set, and each set of sequences shown in two scans.

\underline{\textit{Random Dot Kinematogram}}: 
Random dot kinematograms (RDK) movies were presented as described in \citep{Morrone2000-ov,Wang2023.03.21.533548}, but with different parameters as described below.  
For six scans containing RDK movies, 32 movies (8 flow trajectories $\times$ 2 velocities $\times$ 2 coherencies) were shown ten times per scan.
RDK movie optical flow corresponded to a translational (up/down/left/right), radial (inward / outward w/r/t monitor center), or rotational (clockwise / anticlockwise w/r/t monitor center) trajectory.
RDK movie dots had a velocity of either 0.3 or 0.5 (fraction monitor width / second), and coherency of either 50\% or 100\% with respect to the global optical flow trajectory.
Each dot had a diameter of 1/32 (fraction monitor width, approx. 6.7 degrees at the nearest point) and a lifetime of 1 second.
Each RDK movie was 2 seconds in duration, and movies were randomly assorted into 8 sequences of 4 movies each, for a total of 8 seconds per sequence.  
Three sets of RDK movies were prepared, with each movie unique to its respective set, and each set of RDK movies shown in two scans.

\underline{\textit{Natural Images}}:  Natural image from ImageNet were presented as in \citep{Walker2019-oq, Wang2023.03.21.533548}.  
For six scans containing natural images, 60 images were shown ten times per scan. 
Randomly selected images were center-cropped to 9:16 aspect ratio and converted to gray scale.  
Images were presented for 500 ms, preceded by a 400-600 ms blank gray screen (pixel value 127/255).  
Images were randomly assorted into 6 sequences of 10 images each, for approx. 10 seconds per sequence.
Three sets of natural images were prepared, with each natural image unique to its respective set, and each set of natural images shown in two scans.

\underline{\textit{Gaussian Dots}}: Gaussian dots were presented as in \citep{Wang2023.03.21.533548}, but with different parameters as detailed below.  
For six scans containing gaussian dots, 210 dot presentations (105 positions $\times$ 2 dot intensities) were shown ten times per scan.
Dot positions were drawn from a grid of 15 horizontal (-0.35 to 0.35) by 7 vertical (-0.267 to 0.267) positions, where all positions are reported as fraction of monitor width and 0 is the center of the monitor.  
Dots were presented as either white (pixel value 255 out of 255) or black (pixel value 0) on a gray background (pixel value 127).
Dot standard deviation was 0.07 (fraction monitor width, $ \approx 15  \degree$ at the closest point).
Dot presentations were 300 ms in duration, and were randomly assorted into 6 sequences of 35 dots each, for a total of 10.5 seconds per sequence. 
Because the stimulus was parametrically constructed, the same dots are shown in each of the six scans.
Three sets of gaussian dots that differ in sequence membership and order were prepared, and each set of dots was shown in two scans.

A photodiode (TAOS TSL253) was sealed to the top left corner of the monitor, and the voltage was recorded at 10 kHz and timestamped on the behavior clock (MasterClock PCIe-OSC-HSO-2 card).  
Simultaneous measurement with a luminance meter (LS-100 Konica Minolta) perpendicular to and targeting the center of the monitor was used to generate a lookup table for linear interpolation between photodiode voltage and monitor luminance in cd/m\textsuperscript{2} for 16 equidistant values from 0-255, and one baseline value with the monitor unpowered.

At the beginning of each experimental session, we collected photodiode voltage for 52 full-screen pixel values from 0 to 255 for one second trials.  
The mean photodiode voltage for each trial $V_{pd}$ was fit as a function of the pixel intensity $V_{in}$:

\begin{equation}
     V_{pd} = B + A \times V_{in}^{\gamma}
 \end{equation}

in order to estimate the $\gamma$ value of the monitor ($ \approx 1.60-1.76$).  All stimuli were shown with no $\gamma$ correction.  

During the stimulus presentation, sequence information was encoded in a 3 level signal according to the binary encoding of the flip number assigned in-order.  
This signal underwent a sine convolution, allowing for local peak detection to recover the binary signal.  
The encoded binary signal was reconstructed for >99\% of the flips.  
A linear fit was applied to the trial timestamps in the behavioral and stimulus clocks, and the offset of that fit was applied to the data to align the two clocks, allowing linear interpolation between them. 
The mean photodiode voltage of the sequence encoding signal at pixel values 0 and 255 was used to estimate the luminance range of the monitor during the stimulus, with minimum values between 0.001 and 0.65 cd/m\textsuperscript{2} and maximum values between 8.7 and 11.3 cd/m\textsuperscript{2} in the released scans.  
\subsection*{Preprocessing of neural responses and behavioral data} 

The full two photon imaging processing pipeline is available at (\href{https://github.com/cajal/pipeline}{https://github.com/cajal/pipeline}). 
Raster correction for bidirectional scanning phase row misalignment was performed by iterative greedy search at increasing resolution for the raster phase resulting in the maximum cross-correlation between odd and even rows. 
Motion correction for global tissue movement was performed by shifting each frame in X and Y to maximize the correlation between the cross-power spectra of a single scan frame and a template image, generated from the Gaussian-smoothed average of the Anscombe transform from the middle 2000 frames of the scan. 
Neurons were automatically segmented using constrained non-negative matrix factorization, then detrended and deconvolved to extract estimates of spiking activity, within the CAIMAN pipeline \citep{Giovannucci2019-gw}.  
Cells were further selected by a classifier trained to separate somata versus artifacts based on segmented cell masks, resulting in exclusion of 7.1 - 10.1\% of masks per scan.

Functional and behavioral signals were resampled to 30 Hz by linear spline interpolation.
The mirror motor coordinates of the centroid of each mask was used to assign anatomical coordinates relative to each other and the experimenter's estimate of the pial surface.  
Notably, centroid positional coordinates do not carry information about position relative to the area boundaries, or relative to neurons in other scans.

\paragraph{Representation/Core}

We based our work on the models of \citet{Lurz2020-ua,Franke2022, Ecker2018, sinz2018stimulus, hoefling2022chromatic}, which are able to predict the responses of a large population of mouse V1 neurons with high accuracy. \\
For the GRU baseline, we used rotation-equivariant core from \cite{Ecker2018} with 8 rotations, 8 channels, and 4 layers. The spatial kernels were $9 \times 9$, followed by $7 \times 7$. The GRU module, inspired by \citet{sinz2018stimulus}, was after the core. It had 64 channels (8 channels $\times$ 8 rotations = 64), and both input and recurrent kernels were 9 $\times$ 9. \\
For the 3D Factorized baseline, we used the core inspired by \citet{hoefling2022chromatic} with 3 layers (32, 64, and 128 channels per layer, resp.). The spatial kernels were $11 \times 11$ in the 1st layer and $5 \times 5$ in all of the subsequent layers. Similarly, the temporal kernels were $11 \times 1$ in the 1st layer and $5 \times 1$ afterwards. 

The Ensembled baseline cores were same as for the 3D Factorized baseline.

\paragraph{{Readout}}
To get the scalar neuronal firing rate for each neuron, we computed a linear regression between the core output tensor of dimensions $\mathbf{x}\in \mathbb{R}^{w \times h \times c}$ (\textbf{w}idth, \textbf{h}eight, \textbf{c}hannels) and the linear weight tensor $\mathbf{w} \in \mathbb{R}^{c \times w \times h}$, followed by an ELU offset by one (ELU+1), to keep the response positive. We made use of the recently proposed Gaussian readout \citep{Lurz2020-ua}, which simplifies the regression problem considerably.
The Gaussian readout learns the parameters of a 2D Gaussian distribution $\mathcal{N}(\mu_n, \Sigma_n)$. The mean $\mu_n$ in the readout feature space thus represents the center of a neuron's receptive field in image space, whereas $\Sigma_n$ refers to the uncertainty of the receptive field position. During training, a location of height and width in the core output tensor in each training step is sampled, for every image and neuron. Given a large enough initial $\Sigma_n$ to ensure gradient flow, the uncertainty about the readout location $\Sigma_n$ is decreasing during training, showing that the estimates of the mean location $\mu_n$ becomes more and more reliable.  At inference time (i.e. when evaluating our model), we set the readout to be deterministic and to use the fixed position $\mu_n$. In parallel to learning the position, we learned the weights of the weight tensor of the linear regression of size $c$ per neuron. To learn the positions $\mu_n$, we made use of the retinotopic organization of V1 by coupling the recorded cortical 2d-coordinates $\mathbf{p}_n\in \mathbb R^2$ of each neuron with the estimation of the receptive field position $\mu_n$ of the readout. We achieved this by learning the common function $\mu_n=f(\mathbf{p}_n)$, a randomly initialized linear fully connected MLP of size 2-30-2, shared by all neurons.

\paragraph{{Shifter network}}
We employed a free viewing paradigm when presenting the visual stimuli to the head-fixed mice. Thus, the RF positions of the neurons with respect to the presented images had considerable trial-to-trial variability following any eye movements. We informed our model of the trial dependent shift of neuronal receptive fields due to eye movement by shifting $\mu_n$, the model neuron's receptive field center, using the estimated eye position (see section Neurophysiological experiments above for details of estimating the pupil center). We passed the estimated pupil center through an MLP (the shifter network), a three layer fully connected network with $n=5$ hidden features, followed by a $tanh$ nonlinearity, that calculates the shift in $\Delta$x and $\Delta$y of the neurons receptive field in each trial. We then added this shift to the $\mu_n$ of each neuron.

\paragraph{{Input of behavioral parameters}}
During each presentation of a video, the pupil size and the running speed of the mouse was recorded. We do not have instantaneous pupil dilation change as the target (video) frequency rate is more then the pupil camera sampling frequency. We have used these behavioral parameters to improve the model's predictivity. Because these behavioral parameters have nonlinear modulatory effects, we decided to append them as separate frames to the input images as new channels \citep{Franke2022}, such that each new channel simply consisted of the scalar for the respective behavioral parameter recorded in a particular trial, transformed into stimulus dimension. This enabled the model to predict neural responses as a function of both visual input and behavior. 

\subsection*{Model training}
Both train and validation sets contain only unique videos. We isotropically downsampled all videos to a resolution of $36 \times 64$ px ($h \times w$) per frame. 
Furthermore, we normalized input videos as well as standardized behavioral traces and the target neuronal activities, using the statistics of the training trials of each recording. 
After this we subsampled 150 subsequent frames randomly from each video and trained our network using the batch size $= 8$. 
We used only five competition mice for training, ignoring the pretraining set. 
A gradient update was performed after 5 batches, 1 per mouse. Then, we trained our networks with the training set by minimizing the Poisson loss $\frac{1}{m}\sum_{i=1}^{m}\big(\hat{r}^{(i)}-r^{(i)}\log{\hat{r}^{(i)}}\big)$, where $m$ denotes the number of neurons, $\hat r$ the predicted neuronal response and $r$ the observed response. For Poisson loss each frame was treated independently, and no time component was included. After each epoch, i.e. full pass through the training set, we calculated the correlation between predicted and measured neuronal responses on the validation set and averaged it across all neurons. If the correlation failed to increase for five consecutive epochs, we stopped the training and restored the model to its state after the best performing epoch. 
Then, we either decreased the learning rate by a factor of $0.3$ or stopped training altogether, if the number of learning-rate decay steps was reached (n=4 decay steps). 
We optimized the network's parameters using the Adam optimizer \citep{kingma2014adam}. 
All parameters and hyper-parameters regarding model architecture and training procedure can be found in our \code{sensorium} repository (see Code Availability).

\subsection*{Metrics}
\label{sec:Metrics}
We chose correlation to evaluate the models performance.\\
Since \textbf{correlation} is invariant to shift and scale of the predictions, it does not reward a correct prediction of the absolute value of the neural response but rather the neuron's relative response changes. 
It is bound to $[-1, 1]$ and thus easily interpretable. 
However, without accounting for the unexplainable noise in neural responses, the upper bound of $1$ cannot be reached, which can be misleading. 

\textbf{Single Trial Correlation}
To evaluate model performance on variation between individual trials, we will compute correlation $\rho_{\textrm{st}}$ between predicted single-trial activity $o_{ij}$ and single-trial neuronal responses $r_{ij}$, as

\begin{equation}
    \rho_{\textrm{st}} = \textrm{corr}(\mathbf{r}_{\textrm{st}}, \mathbf{o}_{\textrm{st}}) = \frac{\sum_{i,j}(r_{ij} - \bar{r})(o_{ij} - \bar{o})}{\sqrt{\sum_{i,j}(r_{ij} - \bar{r})^2 \sum_{i,j}(o_{ij} - \bar{o})^2}},
\end{equation}

where $r_{ij}$ is the $i$-th frame of $j$-th video repeat, $o_{ij}$ is the corresponding prediction, $\bar{r}$ is the average response to all the videos in the test subset across all repeats, and $\bar{o}$ is the average prediction for all the videos in the test subset across all repeats.
$\rho_{\textrm{st}}$ is computed independently per neuron and then averaged across all neurons to produce the final metric. 

 


\textbf{Correlation to Average}
We calculate the \textit{correlation to average} $\rho_{\textrm{ta}}$ in a similar way to the single-trial correlation, but we first average the responses and predictions per frame across all video repeats before computing.  

\begin{equation}
    \rho_{\textrm{ta}} = \textrm{corr}(\mathbf{r}_{\textrm{ta}}, \mathbf{o}_{\textrm{ta}}) = \frac{\sum_{i, j}(\bar{r}_{i} - \bar{r})(\bar{o}_{i} - \bar{o})}{\sqrt{\sum_{i}(\bar{r}_{i} - \bar{r})^2 \sum_{i}(\bar{o}_{i} - \bar{o})^2}},
\end{equation}

where $\bar{r}_{i}$ is a response averaged over stimulus repeats for a fixed neuron. 

 


\subsection*{Code and data availability}
Our competition website can be reached under \url{https://www.sensorium-competition.net/}. 
The pretraining dataset split is available for download via DeepLake \citep{hambardzumyan2022deep} upon the competition start, under license CC BY-NC-ND 4.0.
Our coding framework uses general tools including PyTorch, Numpy, scikit-image, matplotlib, seaborn, DataJoint, Jupyter, Docker, CAIMAN, DeepLabCut, Psychtoolbox, Scanimage, and Kubernetes. 
We also used the following custom libraries and code: \code{neuralpredictors} (\url{https://github.com/sinzlab/neuralpredictors}) for torch-based custom functions for model implementation and \code{sensorium} for utilities (\url{https://github.com/ecker-lab/sensorium_2023}).



\section*{References}
\bibliographystyle{apa-good}
\bibliography{reference}
\onecolumn
\newpage


\renewcommand{\figurename}{Supplemental Fig.}
\setcounter{figure}{0}




\begin{figure*}[h!]
\centering
\includegraphics[width=0.95\linewidth]{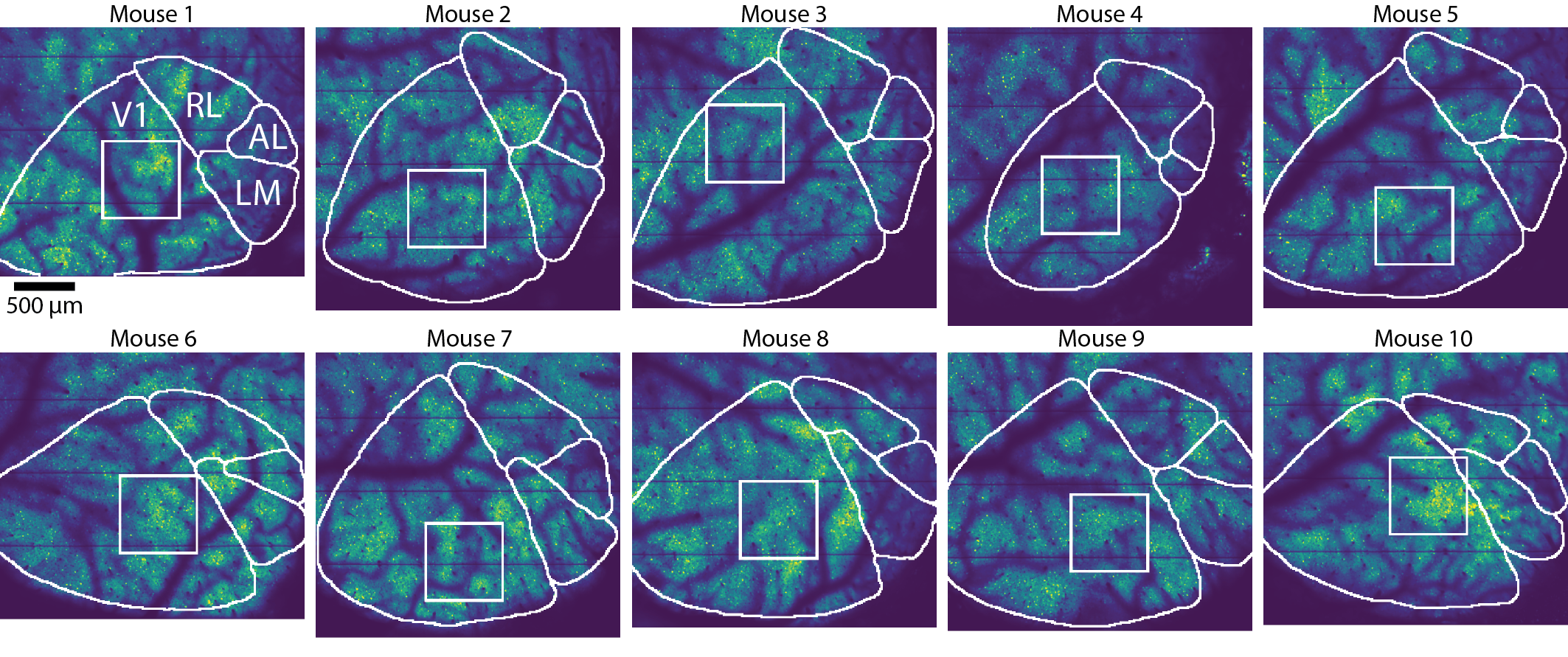}
\caption{\textbf{Scan field locations within primary visual cortex.}
Registered location of the scan field (white rectangle) to retinotopic mapping scan with manually annotated area boundaries (white lines) for primary visual cortex (V1) and anterolateral (AL), rostrolateral (RL), and lateromedial (LM) higher visual areas.
\label{fig:imaging_site_selection}
}
\end{figure*}
\newpage

\end{document}